\documentclass[14pt, fleqn]{extarticle}

\usepackage{latexsym}
\usepackage{amsmath}
\usepackage[colorlinks = true,
            linkcolor = blue,
            urlcolor  = blue,
            citecolor = red,
            anchorcolor = blue]{hyperref}
\usepackage{amsfonts}
\usepackage{amssymb}
\usepackage{graphicx}
\usepackage{color}
\usepackage{tikz}
\usepgflibrary{arrows}
\usepackage{mathbbol}
\usepackage{nicefrac, xfrac}
\usepackage[a4paper,bindingoffset=0.2in,
           left=1in,right=1in,top=1in,bottom=1in]{geometry}

\title{Classical and quantum particles\\ from nongeneric conformal orbits}
\author{
Piotr Kosi\'nski\footnote {piotr.kosinski@uni.lodz.pl}, Pawe\l\, Ma\'slanka \footnote{pawel.maslanka@uni.lodz.pl}\\
\small \textit {Faculty of Physics and Applied Informatics}\\
\small \textit {University of Lodz, Poland}\\
}

\date{}
\begin{document}
\maketitle
\begin{abstract}
\par The nongeneric six- and eightdimensional orbits of SO(4,2) are described in explicitly covariant way. The relevant Hamiltonian dynamical systems are constructed and canonically quantized. It is shown that the resulting unitary representations of conformal group fit into the classification described by Mack (G. Mack, Comm. Math. Phys. \underline{55} (1977), 1).
\end{abstract}

\section{Introduction} 
\label{I}
\par Looking into any textbook on quantum mechanics one finds the following general prescription for constructing a consistent quantum theory. One starts with a classical dynamical system which can be put in Hamiltonian form. Once this is done the algorithm of canonical quantization is applied which consists in replacing classical dynamical variables (i. e. functions on classical phase space) by selfadjoint operators acting in some Hilbert space of states (which is a quantum counterpart of phase space). This replacement should be made in a way to preserve as many algebraic relations as possible while the Poisson bracket is replaced by the commutator divided by $i\hslash$. It is well known \cite{b1} that such an ambitious program cannot be performed as it follows from Groenewold-Van Hove theorem. The latter states that there exists no "quantization map" Q mapping the classical dynamical variables (i.e. functions on the phase space) to the operators in Hilbert space which: \emph{(i)} sends $1$ to the identity operator; classical canonical variables $(x_{i}, p_{i})$ to their standard quantum counterparts $\hat{x}_{i}, \hat{p}_{i}$; any polynomial in $x_{i}, p_{i}$ to some (appropriately ordered) polynomial in $\hat{x}_{i}, \hat{p}_{i}$, and \emph{(ii)} transforms the Poisson bracket into the commutator divided by $i\hslash$. Luckily enough, not all dynamical variables are equally interesting from physical point of view. We are mainly interested in those following (via Noether theorem) from or related to the symmetries exhibited by the dynamics under consideration. If this more modest point of view is accepted the situation improves considerably and canonical quantization yields consistent structure. It should be, however, stressed that there are some important exceptions where certain symmetries do not survive quantization. If this happens to be the case, quantum theory may appear to be internally inconsistent (this concerns gauge theories) or some selection rules are broken; the typical example of the latter situation is the decay $\pi^{o}\rightarrow \gamma+\gamma$.
\par Putting aside the troubles related to Groenewold-Van Hove theorem one encounters the more basic problem. Canonical quantization yields quantum theory which in the limit $\hslash\rightarrow 0$ describes the classical system we have started with. However, when considering the world on its basic microscopic level it seems to make no sense to speak about some classical theory as a starting point. On the contrary, one should start with quantum theory inferred from some basic first principles; the macroscopic classical world becomes then the derived concept. 
\par From practical point of view the above first principles are usually related to the postulated symmetries of elementary interactions. Given a (space-time) symmetry group one defines elementary systems (particles) as those described by irreducible representations of the group under consideration. The main observables are then identified (via Noether theorem) with generators and elements of enveloping algebra of the relevant Lie algebra. Once all irreducible unitary representations are classified and constructed, the elementary dynamical systems (particles) are fully described. Then one can construct multiparticle states by taking the tensor products of irreducible representations. However, the resulting formalism describes noninteracting particles. Interactions are introduced by an appropriate modification of generators acting in tensor product spaces. The way this can be achieved depends on the structure of the relevant symmetry group/algebra. For example, it is quite easy for the case of Galilean symmetry since the generator of time translations (Hamiltonian) does not appear on the right hand side of Galilei algebra. On the contrary, in the case of Poincar\'{e} symmetry introducing interaction converts theory into quantum field theory as nicely explained in Weinberg's monograph \cite{b2}.
\par The approach based on some first (symmetry) principles seems to dominate conceptually the one addressing to naive quantization of classical dynamics. However, it appears that the situation is more subtle and interesting. It has been realized that the unitary representations of many Lie groups can be obtained by considering certain symplectic manifolds and applying the algorithm closely resembling quantization procedure. This is the core of the so called orbit method developed by Kirillov, Kostant and Souriau \cite{b3,b4,b5,b6,b7,b8}. The basic idea is the observation that the coadjoint orbits of Lie groups are equipped with invariant (under the coadjoint action of the group under consideration) symplectic structure which can be easily recovered from the Maurer-Cartan form \cite{b9}, \cite{b10}. Once this invariant symplectic structure is explicitly described one can apply the procedure, which can be vaguely called canonical quantization, to find an unitary representation of the group considered. This works perfectly for solvable groups while in the case of compact and non-compact semisimple ones one is faced with some obstacles and troubles \cite{b6}.
\par The existence of Kirillov-Kostant-Souriau formalism suggests that the approach consisting in "canonical quantization" of classical Hamiltonian dynamics is, in fact, not less general than the one starting from first (symmetry) principles. The mathematical basis for such a point of view is provided by the following statement \cite{b10}: given a Lie group $G$, all symplectic manifolds on which $G$ acts transitively preserving symplectic structure are (up to topological subtleties like taking the covering etc.) coadjoint orbits of $G$. Assume that $G$ involves space-time transformations; then the generator of time translations (the Hamiltonian) belongs to the Lie algebra of $G$. If we further assume that the dynamical system is elementary, i.e. $G$ acts transitively on the phase space, then the classification of $G$-invariant classical Hamiltonian systems reduces to the classification of coadjoint orbits of $G$.
\par The resulting classical systems are then canonically quantized yielding the quantum ones equivalent to those obtained from "first principles".
\par This philosophy, in more or less direct form, has been already applied to relativistic systems exhibiting Poincar\'{e} symmetry \cite{b11,b12,b13,b14,b17,b18,b19,b21,b23,b24,b25,b26,b27,b28,b30,b31}. In the present paper we consider the orbit method in the context of physically important case of conformal symmetry in fourdimensional space-time. The conformal transformations form the noncompact semisimple group locally isomorphic to $SO(4,2)\,(\sim SU(2,2)/\mathbb{Z}_{4})$. Its coadjoint orbits can be classified using, for example, twistor formalism \cite{b32}. The generic ones are 12-dimensional; there exist also nongeneric orbits of dimensions 10, 8 and 6 (and, obviously, the trivial one). We will be interested here in 6- and 8-dimensional orbits. The former describe massless particles of definite helicity and were already discussed in the framework of orbit method (cf. Refs. \cite{b7, b15, b16, b33}); here we provide explicitly covariant approach to 6-dimensional case. Further, we show that our 8-dimensional orbits describe Hamiltonian mechanics of spinless particles of arbitrary mass. Canonical quantization is directly performed leading to the quantum description based on unitary irreducible representations of conformal group.
\par Let us note that all unitary representations of conformal group with positive energy were classified by Mack \cite{b34} (see also references therein). We show that our representations fit perfectly into Mack scheme. In the subsequent paper we will extend our straightforward quantization method to other orbits in order to recover the remaining representations classified by Mack and describe them more directly in physical terms. The important point here seems to be that the positivity of energy condition seems to exclude representations for which the application of the orbit method may lead to some troubles mentioned in Ref. \cite{b6}. It has to be, however, noted that even in the relatively simple case considered here certain representations, corresponding to some values of the spectral parameter characterizing the ground state of conformal Hamiltonian, cannot be recovered by the method of orbits. 

\section{Sixdimensional orbits} 
\label{II}

\par We consider sixdimensional pseudoeuclidean space equipped with the metric tensor
\setlength{\jot}{15pt}
\begin {align} 
\label {al1}
g_{ab}=g^{ab}=diag (+----+)
\end {align}
and $a,b=0,1,2,3,5,6$; we adopt here the Todorov \cite{b32} convention omitting the index $4$. 
\par $SO(4,2)$ group acts on our space by pseudoorthogonal transformations
\setlength{\jot}{15pt}
\begin {align} 
\label {al2}
&\xi^{a'}=D^{a}_{\phantom{a}b}(g)\xi^{b}\\ \nonumber
&g_{ab}D^{a}_{\phantom{a}c}(g)D^{b}_{\phantom{b}d}(g)=g_{cd}
\end {align}
Denoting by $X_{ab}=-X_{ba}$ the generators of the Lie algebra of $SO(4,2)$ one finds the following commutation rules
\setlength{\jot}{15pt}
\begin {align} 
\label {al3}
[X_{ab},X_{cd}]=i(g_{ad}X_{bc}+g_{bc}X_{ad}-g_{ac}X_{bd}-g_{bd}X_{ac})
\end {align}
Let $\{\tilde{X}^{ab}\}$ be the basis in the dual space to our Lie algebra; any element $\tilde{X}$ of dual space can be written as
\setlength{\jot}{15pt}
\begin {align} 
\label {al4}
\tilde{X}=\zeta_{ab}\tilde{X}^{ab}\quad\textbf{,}\quad \zeta_{ab}=-\zeta_{ba}
\end {align}
The coadjoint action of $SO(4,2)$ reads
\setlength{\jot}{15pt}
\begin {align} 
\label {al5}
\zeta'_{ab}=D^{c}_{\phantom{c}a}(g^{-1})D^{d}_{\phantom{d}b}(g^{-1})\zeta_{cd}
\end {align}
The Poisson structure can be inferred directly from eq. (\ref{al3}):
\setlength{\jot}{15pt}
\begin {align} 
\label {al6}
\{\zeta_{ab},\zeta_{cd}\}=g_{ad}\zeta_{bc}+g_{bc}\zeta_{ad}-g_{ac}\zeta_{bd}-g_{bd}\zeta_{ac}\;\text{;}
\end {align}
it is invariant under the transformations (\ref{al5}). Moreover, the latter are generated by $\zeta_{ab}$: the infinitesimal transformations are obtained by taking the appropriate Poisson bracket.
\par The Poisson structure (\ref{al6}) is degenerate. There exist three functionally independent Casimir functions Poisson-commuting with all $\zeta_{ab}$:
\setlength{\jot}{15pt}
\begin {align} 
\label {al7}
&C_{2}\equiv \zeta_{ab}\zeta^{ab}\\ \nonumber
&C_{3}\equiv \epsilon^{abcdef}\zeta_{ab}\zeta_{cd}\zeta_{ef}\;(\epsilon^{012356}=1)\\ \nonumber
&C_{4}\equiv \zeta_{ab}\zeta^{b}_{\phantom{b}c}\zeta^{c}_{\phantom{c}d}\zeta^{da}
\end {align}
where raising/lowering indices is performed with the help of metric tensor(s).
\par Let us remind the relations between the generators $\zeta_{ab}$ of (pseudo)rotations in sixdimensional pseudoeuclidean space and the physical generators of conformal group \cite{b35}:
\setlength{\jot}{15pt}
\begin {align} 
\label {al8}
p_{0}&\equiv \zeta_{05}+\zeta_{06}\\ \nonumber
p_{i}&\equiv \zeta_{i5}+\zeta_{i6}\quad \text{,}\quad i=1,2,3\\ \nonumber
m_{ij}&\equiv\zeta_{ij}\quad \text{,}\quad i,j=1,2,3\\ \nonumber
m_{0i}&\equiv\zeta_{0i}\quad \text{,}\quad i=1,2,3\\ \nonumber
d&=\zeta_{56}\\ \nonumber
k_{0}&=-\zeta_{05}+\zeta_{06}\\ \nonumber
k_{i}&=-\zeta_{i5}+\zeta_{i6}\quad \text{;}\quad i=1,2,3
\end {align}
here $(p_{0},p_{i})$, $(m_{0i},m_{ij})$, $d$, $(k_{0},k_{i})$ are the generators of translations, Lorentz subgroup, dilatations and special conformal transformations, respectively.
\par Now, due to the existence of three functionally independent Casimir functions, the generic orbits are 12-dimensional. However, as mentioned above, there exist also nongeneric, lower dimensional ones. They can be obtained by imposing further $SO(4,2)$-covariant constraints. Since the (infinitesimal) coadjoint action is obtained by taking the appropriate Poisson bracket, these constraints generate the ideal in the Poisson algebra of functions on dual space. Therefore, the Poisson bracket is consistently defined on the submanifold described by covariant constraints; if the submanifold is an orbit, the Poisson bracket is nondegenerate. 
\par In order to define a simple nongeneric orbit we impose the following constraints:
\setlength{\jot}{15pt}
\begin {align} 
\label {al9}
\zeta_{a}^{\phantom{a}c}\zeta_{cb}+\rho g_{ab}=0\quad \text{,}\quad \rho\neq0
\end {align}
Let us denote
\setlength{\jot}{15pt}
\begin {align} 
\label {al10}
\zeta_{0\mu}\equiv\omega_{\mu}\quad \text{,}\quad \zeta_{6\mu}\equiv z_{\mu}\quad \text{,}\quad \mu=1,2,3,5
\end {align}
By considering various combinations of indices $a,b$ in eq. (\ref{al9}) one easily finds:
\setlength{\jot}{15pt}
\begin {align} 
\label {al11}
\omega_{\mu}z_{\mu}=0\quad \text{,}\quad \omega_{\mu}\omega_{\mu}=z_{\mu}z_{\mu}
\setlength{\jot}{15pt}
\end {align}
\begin {align} 
\label {al12}
\zeta_{06}=\sqrt{\omega_{\mu}\omega_{\mu}+\rho}=\sqrt{z_{\mu}z_{\mu}+\rho}\quad \text{,}\quad \rho \geq 0
\end {align}
\setlength{\jot}{15pt}
\begin {align} 
\label {al13}
\zeta_{\mu \nu}=\frac{\zeta_{06}}{\zeta^{2}_{06}-\rho}(\omega_{\mu}z_{\nu}-\omega_{\nu}z_{\mu})\pm
\frac{\sqrt{\rho}}{\zeta^{2}_{06}-\rho}\epsilon_{\mu\nu\alpha\beta}\omega_{\alpha}z_{\beta}\quad\quad(\epsilon_{1235}=1)\;\;\text{.}
\end {align}
Therefore, our submanifold is characterized by two fourdimensional euclidean vectors which are orthogonal and have the same length; it is sixdimensional. The lowest dimension of a nontrivial orbit is also six so the submanifold under consideration is at most discrete sum of orbits. Actually, it is the sum of two orbits corresponding to two choices of sign in front of $\sqrt{\rho}$ in eq. (\ref{al13}). To see this let us consider the "$+$" sign as an example. By $SO(4)$ rotations generated by $\zeta_{\mu \nu}$, $\mu, \nu = 1,2,3,5$, one can set $\omega_{\mu}=(\omega,0,0,0)$, $z_{\mu}=(0,\omega,0,0)$. Then, from eq. (\ref{al13}), we find 
\setlength{\jot}{15pt}
\begin {align}
\label {al14}
\zeta_{12}=-\zeta_{21}=\zeta_{06}\quad \text{,}\quad \zeta_{35}=-\zeta_{53}=\sqrt{\rho}
\end {align}
and $\zeta_{\mu\nu}=0$ otherwise. Now, by applying the Lorentz boost generated by $\zeta_{16}$ one gets
\setlength{\jot}{15pt}
\begin {align}
\label {al15}
\zeta_{01}(\lambda)&=\zeta_{01}ch\lambda+\zeta_{06}sh\lambda\\ \nonumber
\zeta_{06}(\lambda)&=\zeta_{01}sh\lambda+\zeta_{06}ch\lambda\\ \nonumber
\zeta_{12}(\lambda)&=\zeta_{12}ch\lambda+\zeta_{62}sh\lambda\\ \nonumber
\zeta_{62}(\lambda)&=\zeta_{12}sh\lambda+\zeta_{62}ch\lambda
\end {align}
with other parameters remaining unchanged. Let us first assume $\rho>0$; then $\omega<\zeta_{06}$. Putting $th\lambda=-\frac{\omega}{\zeta_{06}} (\vert\omega\vert<\zeta_{06})$ one finds $\zeta_{01}=0$, $\zeta_{06}=\sqrt{\rho}$, $\zeta_{12}=\sqrt{\rho}$, $\zeta_{62}=0$. Therefore, any point of our submanifold can be transformed to the canonical one: $\zeta_{12}=-\zeta_{21}=\zeta_{06}=-\zeta_{60}=\zeta_{35}=-\zeta_{53}=\sqrt{\rho}$, $\zeta_{ab}=0$ otherwise. Similar reasoning applies to the choice of "$-$" sign in eq. (\ref{al13}). We conclude that our submanifold defined by eq. (\ref{al9}) consists of two connected components, each being the coadjoint orbit. In what follows we describe the orbit corresponding to the choice $\zeta_{35}=\sqrt{\rho}$; the second one is obtained by making the replacement $\sqrt{\rho}\rightarrow-\sqrt{\rho}$. For $\rho=0$ we get $\zeta_{06}=\zeta_{01}=\zeta_{12}=\zeta_{62}$, so we can use eqs. (\ref{al15}) to put all of them equal to one; again we obtain an orbit with canonical point $\zeta_{06}=\zeta_{01}=\zeta_{12}=\zeta_{62}=1$; $\zeta_{ab}=0$ otherwise.
\par The basic Poisson brackets can be immediately read off from eqs. (\ref{al6}) and (\ref{al10})-(\ref{al13}):
\setlength{\jot}{15pt}
\begin {align}
\label {al16}
\{\omega_{\mu},\omega_{\nu}\}=-\frac{\sqrt{\omega_{\alpha}\omega_{\alpha}+\rho}}{\omega_{\alpha}\omega_{\alpha}}\;(\omega_{\mu}z_{\nu}-\omega_{\nu}z_{\mu})-\frac{\sqrt{\rho}}{\omega_{\alpha}\omega_{\alpha}}\;\epsilon_{\mu\nu\alpha\beta}\,\omega_{\alpha}z_{\beta}
\end {align}
\setlength{\jot}{15pt}
\begin {align}
\label {al17}
\{z_{\mu},z_{\nu}\}=-\frac{\sqrt{z_{\alpha}z_{\alpha}+\rho}}{z_{\alpha}z_{\alpha}}\;(\omega_{\mu}z_{\nu}-\omega_{\nu}z_{\mu})-\frac{\sqrt{\rho}}{z_{\alpha}z_{\alpha}}\;\epsilon_{\mu\nu\alpha\beta}\,\omega_{\alpha}z_{\beta}
\end {align}
\setlength{\jot}{15pt}
\begin {align}
\label {al18}
\{\omega_{\mu},z_{\nu}\}=\sqrt{\omega_{\alpha}\omega_{\alpha}+\rho}\;\;\delta_{\mu\nu}\equiv \sqrt{z_{\alpha}z_{\alpha}+\rho}\;\;\delta_{\mu \nu}
\end {align}
Eqs. (\ref{al11})-(\ref{al13}) and (\ref{al16})-(\ref{al18}) provide the complete description of our sixdimensional orbit as symplectic manifold. For "physical" generators (\ref{al8}) one finds
\setlength{\jot}{15pt}
\begin {align}
\label {al19}
p_{0}&=\omega_{5}+\sqrt{\omega_{\alpha}\omega_{\alpha}+\rho}\\ \nonumber
p_{i}&=\frac{\sqrt{\omega_{\alpha}\omega_{\alpha}+\rho}}{\omega_{\alpha}\omega_{\alpha}} z_{5}\omega_{i}-\Bigg(\frac{\sqrt{\omega_{\alpha}\omega_{\alpha}+\rho}}{\omega_{\alpha}\omega_{\alpha}}\omega_{5}+1\Bigg)z_{i}+\frac{\sqrt{\rho}}{\omega_{\alpha}\omega_{\alpha}}\epsilon_{ijk}\,\omega_{j}z_{k}\\ \nonumber
k_{0}&=-\omega_{5}+\sqrt{\omega_{\alpha}\omega_{\alpha}+\rho}\\ \nonumber
k_{i}&=-\frac{\sqrt{\omega_{\alpha}\omega_{\alpha}+\rho}}{\omega_{\alpha}\omega_{\alpha}} z_{5}\omega_{i}+\Bigg(\frac{\sqrt{\omega_{\alpha}\omega_{\alpha}+\rho}}{\omega_{\alpha}\omega_{\alpha}}\omega_{5}+1\Bigg)z_{i}-\frac{\sqrt{\rho}}{\omega_{\alpha}\omega_{\alpha}}\epsilon_{ijk}\,\omega_{j}z_{k}\\ \nonumber
d&=-z_{5}\\ \nonumber
m_{0i}&=\omega_{i} \\ \nonumber
m_{ij}&=\frac{\sqrt{\omega_{\alpha}\omega_{\alpha}+\rho}}{\omega_{\alpha}\omega_{\alpha}} (\omega_{i}z_{j}-\omega_{j}z_{i})+\frac{\sqrt{\rho}}{\omega_{\alpha}\omega_{\alpha}}\epsilon_{ijk}(z_{5}\omega_{k}-\omega_{5}z_{k})
\end {align}
One can check that $p_{0}^{\phantom{0}2}-\vec{p}\,^{2}=0$, i.e. we are dealing with massless particles. Moreover, the Poisson brackets (\ref{al16})-(\ref{al18}) imply that the generators (\ref{al19}) obey the conformal algebra with respect to Poisson brackets.
\par The above formulae look quite complicated. However, we are interested in finding representation of generators in terms of Darboux coordinates. One can try to take the components $p_{i}$ of the threemomentum as half of them. Moreover, we define
\setlength{\jot}{15pt}
\begin {align}
\label {al20}
x_{i}\equiv -\frac{\omega_{i}}{\omega_{5}+\sqrt{\omega_{\alpha}\omega_{\alpha}+\rho}}\equiv\frac{m_{i0}}{p_{0}}
\end {align}
The six independent variables $\vec{x}$ and $\vec{p}$ obey
\setlength{\jot}{15pt}
\begin {align}
\label {al21}
\{x_{i},x_{j}\}=-\frac{\sqrt{\rho}}{p_{0}^{\phantom{0}3}}\epsilon_{ijk}p_{k}
\end {align}
\setlength{\jot}{15pt}
\begin {align}
\label {al22}
\{x_{i},p_{j}\}=\delta_{ij}
\end {align}
\setlength{\jot}{15pt}
\begin {align}
\label {al23}
\{p_{i},p_{j}\}=0
\end {align}
Eq. (\ref{al21}) tells us that we are dealing with monopole in momentum space \cite{b20,b22,b36,b29}. This implies that the Darboux coordinates can be introduced only locally. However, the generators of conformal group can be conveniently expressed in terms of $\vec{x}$ and $\vec{p}$:
\setlength{\jot}{15pt}
\begin {align}
\label {al24}
p_{0}&=\vert \vec{p}\,\vert\\ \nonumber
d&=x_{k}p_{k}\\ \nonumber
m_{0i}&=-p_{0}x_{i}\\ \nonumber
m_{ij}&=x_{i}p_{j}-x_{j}p_{i}+\frac{\sqrt{\rho}}{p_{0}}\epsilon_{ijk}\,p_{k}\\ \nonumber
k_{0}&=p_{0}\vec{x}\,^{2}+\frac{\rho}{p_{0}}\\ \nonumber
k_{i}&=\vec{x}\,^{2}p_{i}-2p_{k}x_{k}x_{i}-\frac{\rho p_{i}}{p_{0}^{\phantom{0}2}}-\frac{2\sqrt{\rho}}{p_{0}}\epsilon_{ijk}\,x_{j}p_{k}
\end {align}
The above generators describe the classical massless particles with helicity $\sqrt{\rho}$.
\par The quantum version of the above dynamics can be obtained by applying the canonical quantization procedure. It appears that the ordering problem for the generators is slightly more complicated than in the case of Poincar\'{e} group. There appear some additional, $\hslash$-dependent corrections. Once they are correctly recognized one obtains the generators obeying the commutation rules. In order to find their explicit form note that $p_{i}$'s form the maximal set of commuting operators. Therefore, they can be chosen as multiplication operators. Then $x_{i}$'s become the covariant derivatives for the monopole field configuration in momentum space. On the global level the quantization is consistent only provided the charge (helicity) $\sqrt{\rho}$ is appropriately quantized. 
\par This quantization procedure is described in some detail in Ref. \cite{b33}) and won't be discussed here. As the final result we obtain the infinitesimal form of unitary positive energy representation of the conformal group coinciding with the case (\ref{al5}) in Mack's terminology: $m=0$, $j_{1}j_{2}=0$, $\delta =j_{1}+j_{2}+1$.
\section{Eightdimensional orbits} 
\label{III}
\par Consider the submanifold of dual space defined by the equations:
\setlength{\jot}{15pt}
\begin {align}
\label {al25}
\epsilon^{abcdef}\zeta_{cd}\zeta_{ef}=0
\end {align}
\setlength{\jot}{15pt}
\begin {align}
\label {al26}
C_{2}\equiv \zeta_{ab}\zeta^{ab}=2\lambda^{2}\quad \text{,}\quad \lambda>0
\end {align}
They can be solved as follows. Assuming, as previously, that the Greek indices take the values $1,2,3,5$ and putting again
\setlength{\jot}{15pt}
\begin {align}
\label {al27}
\zeta_{0\mu}\equiv\omega_{\mu}\quad \text{,}\quad \zeta_{6\mu}\equiv z_{\mu}
\end {align}
we find from eq. (\ref{al25})
\setlength{\jot}{15pt}
\begin {align}
\label {al28}
\zeta_{\mu\nu}=\frac{1}{\zeta_{06}}(\omega_{\mu}z_{\nu}-\omega_{\nu}z_{\mu})
\end {align}
It remains to determine $\zeta_{06}$. Inserting (\ref{al27}) and (\ref{al28}) into eq. (\ref{al26}) one finds biquadratic equation for $\zeta_{06}$:
\setlength{\jot}{15pt}
\begin {align}
\label {al29}
\zeta_{06}^{4}-(\omega^{2}+z^{2}+\lambda^{2})\zeta_{06}^{2}+\omega^{2}\cdot z^{2}-(\omega\cdot z)^{2}=0
\end {align}
where $\omega^{2}\equiv \omega_{\alpha}\omega_{\alpha}$ etc. In order to provide positive-energy real-mass dynamics (see below) we choose 
\setlength{\jot}{15pt}
\begin {align}
\label {al30}
\zeta_{06}=\Bigg[\frac{1}{2}\bigg(\omega^{2}+z^{2}+\lambda^{2}+\sqrt{(\omega^{2}+z^{2}+\lambda^{2})^{2}-4\Big(\omega^{2}\cdot z^{2}-(w \cdot z)^{2}\Big)}\bigg)\Bigg]^{\frac{1}{2}}
\end {align}
It is easy to check that (\ref{al25}) and (\ref{al26}) do not provide any further constraints. Therefore, our submanifold is given by eqs. (\ref{al27}), (\ref{al28}), (\ref{al30}) in terms of two arbitrary fourvectors $\omega_{\mu}, z_{\mu}$. It is eightdimensional. 
\par In order to show that (\ref{al25}), (\ref{al26}) define a coadjoint orbit let us first note that by $SO(4)$ rotations acting on the antisymmetric tensor $\zeta_{\mu\nu}$ one can achieve
\setlength{\jot}{15pt}
\begin {align}
\label {al31}
\zeta_{23}=0\quad \text{,}\quad \zeta_{13}=0\quad \text{,}\quad \zeta_{15}=0\quad \text{,}\quad \zeta_{25}=0
\end {align}
(this is immediately seen if we realize that locally $SO(4)\sim SO(3)\times SO(3)$). Therefore, only $\zeta_{12}=-\zeta_{21}$, $\zeta_{35}=-\zeta_{53}$ can be nonvanishing. Putting $(ab)=(06)$ in eq. (\ref{al25}) yields $\zeta_{12}\cdot \zeta_{35}=0$. Due to $SO(4)$ symmetry one can assume $\zeta_{35}=0$ without loss of generality, so finally only one component of $\zeta_{\mu \nu}(\mu \nu =1,2,3,5)$, $\zeta_{12}=-\zeta_{21}$, is nonvanishing.\\
Choosing in (\ref{al25}) $(ab)=(0\mu)$ or $(6\mu)$ we get $\omega_{3}=\omega_{5}=0$, $z_{3}=z_{5}=0$. We are left with $\omega_{1}, \omega_{2}, z_{1}, z_{2},\zeta_{06}$ given by eq. (\ref{al30}) and
\setlength{\jot}{15pt}
\begin {align}
\label {al32}
\zeta_{12}=\frac{1}{\zeta_{06}}(\omega_{1}z_{2}-\omega_{2}z_{1})
\end {align}
$\zeta_{12}$ and $\zeta_{06}$ generate rotations in $(12)$ and $(06)$ euclidean planes. This allows to put (without spoiling the remaining constraints) $\omega_{2}=0$, $z_{1}=0$. Eq. (\ref{al30}) implies $\zeta_{06}>\vert \omega_{1}\vert$, $\zeta_{06}>\vert z_{2}\vert$. Using these inequalities one easily shows that the boosts generated by $\zeta_{16}$ and $\zeta_{02}$ allow us to put $\omega_{1}=0$, $z_{2}=0$ (again without spoiling other constraints) so that also $\zeta_{12}=0$ (cf. eq. (\ref{al32})). It remains only one nonvanishing element $\zeta_{06}=-\zeta_{60}=\lambda$. Therefore, eqs. (\ref{al25}) and (\ref{al26}) define the eightdimensional orbit. 
\par The relevant Poisson brackets can be again read off from eq. (\ref{al6})
\setlength{\jot}{15pt}
\begin {align}
\label {al33}
\{\omega_{\mu}, \omega_{\nu}\}&=-\frac{1}{\zeta_{06}}(\omega_{\mu}z_{\nu}-\omega_{\nu}z_{\mu})\\ \nonumber
\{z_{\mu}, z_{\nu}\}&=-\frac{1}{\zeta_{06}}(\omega_{\mu}z_{\nu}-\omega_{\nu}z_{\mu})\\ \nonumber
\{\omega_{\mu}, z_{\nu}\}&=\zeta_{06}\delta_{\mu\nu}
\end {align}
The generators of conformal group, when expressed in terms of $\omega_{\mu}$ and $z_{\mu}$, read
\setlength{\jot}{15pt}
\begin {align}
\label {al34}
p_{0}&=\zeta_{06}+\omega_{5}\\ \nonumber
p_{i}&=\frac{z_{5}}{\zeta_{06}}\omega_{i}-\bigg(\frac{\omega_{5}}{\zeta_{06}}+1\bigg)z_{i}\\ \nonumber
k_{0}&=\zeta_{06}-\omega_{5}\\ \nonumber
k_{i}&=-\frac{z_{5}}{\zeta_{06}}\omega_{i}+\bigg(\frac{\omega_{5}}{\zeta_{06}}-1\bigg)
z_{i}\\ \nonumber
d&=-z_{5}\\ \nonumber
m_{0i}&=\omega_{i}\\ \nonumber
m_{ij}&=\frac{1}{\zeta_{06}}\bigg(\omega_{i}z_{j}-\omega_{j}z_{i}\bigg)
\end {align}
Contrary to the case of sixdimensional orbits we can now find global Darboux coordinates. First, we adopt $p_{i}$ as a part of Darboux variables and define, as previously.
\setlength{\jot}{15pt}
\begin {align}
\label {al35}
x_{i}\equiv -\frac{\omega_{i}}{\omega_{5}+\zeta_{06}}\equiv -\frac{m_{0i}}{p_{0}}
\end {align}
Then
\setlength{\jot}{15pt}
\begin {align}
\label {al36}
\{x_{i}, x_{j}\}=0\\ \nonumber
\{x_{i},p_{j}\}=\delta_{ij}\\ \nonumber
\{p_{i},p_{j}\}=0
\end {align}
We still need one pair of canonically conjugated variables. Let us first note that $p_{0}$ is no longer a function of $\vec{p}$; it is an independent variable. Now, using the definition of $p_{0}$ and eq. (\ref{al30}) one can show that
\setlength{\jot}{15pt}
\begin {align}
\label {al37}
p_{0}^{2}-\vec{p}\,^{2}>0
\end {align}
Therefore, the mass variable defined as
\setlength{\jot}{15pt}
\begin {align}
\label {al38}
m=\sqrt{p_{0}^{2}-\vec{p}\,^{2}}>0
\end {align}
is real positive. Moreover,
\setlength{\jot}{15pt}
\begin {align}
\label {al39}
\{x_{i},m\}&=0\\ \nonumber
\{p_{i},m\}&=0
\end {align}
The final step is to construct the variable canonically conjugated to $m$. We set
\setlength{\jot}{15pt}
\begin {align}
\label {al40}
y\equiv \frac{-z_{5}-x_{k}p_{k}}{m}
\end {align}
where $\vec{x}$, $\vec{p}$, $m$ are already known functions of $\omega_{\mu}$ and $z_{\mu}$. Then
\setlength{\jot}{15pt}
\begin {align}
\label {al41}
\{x_{i},y\}&=0\\ \nonumber
\{p_{i},y\}&=0\\ \nonumber
\{y,m\}&=1
\end {align}
It is not difficult to show that the transformation $(\omega_{\mu},z_{\mu})\rightarrow(\vec{x},\vec{p},y,m)$ is invertible (actually, it follows almost immediately from the formulae written below); therefore, we found Darboux coordinates for the Poisson structure (\ref{al33}).
\par Finally, let us express the conformal generators in terms of Darboux variables. The resulting formulae read: 
\setlength{\jot}{15pt}
\begin {align}
\label {al42}
p_{0}&=\sqrt{\vec{p}\,^{2}+m^{2}} \\ \nonumber
m_{0i}&=-p_{0}x_{i} \\ \nonumber
m_{ij}&=x_{i}p_{j}-x_{j}p_{i}\\ \nonumber
d&=x_{k}p_{k}+y\cdot m\\ \nonumber
k_{0}&=p_{0}\bigg(\frac{\lambda}{m^{2}}+\vec{x}\,^{2}+\vec{y}\,^{2}\bigg)\\ \nonumber
k_{i}&=\frac{k_{0}p_{i}}{p_{0}}-2dx_{i}
\end {align}

Let us note that the expression for $k_{0}$ is equivalent to the biquadratic equation for $\zeta_{06}$ while the one for $k_{i}$ becomes a simple identity once the expressions for $p_{i}$, $p_{0}$, $k_{i}$, $k_{0}$, $d$ and $x_{i}$ in terms of $\zeta_{ab}$ are inserted. It follows from eqs. (\ref{al42}) that our system consists of single spinless particle with variable mass.
\par Let us consider the quantum counterpart of the above dynamics. The problem simplifies due to the fact that there exist global Darboux coordinates. We can apply the "naive" canonical quantization method. Assume we have already found the operators $\vec{X}$, $\vec{P}$, $M$, $Y$ representing the canonical variables (in what follows we use capital letters to denote quantum dynamical variables); this is essentially unique due to the Stone-von Neumann theorem. Then we would like to construct the relevant formulae for the conformal group generators, i.e. the quantum counterpart of eqs. (\ref{al42}). To this end we have to symmetrize appropriately and/or add some $\hslash$-dependent terms in order to fulfil the following conditions:\emph{(i)} the generators should be (formally - see below) selfadjoint; \emph{(ii)} they should obey the commutation rules of the conformal Lie algebra. This can be done by applying the simplest symmetrization procedure and allowing for additional, $\hslash$-dependent correction terms; the general structure of the latter can be quite easily predicted. The final result reads
\setlength{\jot}{15pt}
\begin {align}
\label {al43}
P_{0}&=\sqrt{\vec{P}\,^{2}+M^{2}} \\ \nonumber
M_{ij}&=X_{i}P_{j}-X_{j}P_{i}\\ \nonumber
M_{i0}&=\frac{1}{2}(X_{i}P_{0}+P_{0}X_{i})\\ \nonumber
D&=\frac{1}{2}(X_{k}P_{k}+P_{k}X_{k}+YM+MY)\\ \nonumber
K_{0}&=\frac{1}{2}\Bigg[P_{0}\bigg(\frac{\lambda^{2}}{M^{2}}+\vec{X}\,^{2}+Y^{2}\bigg)
+\bigg(\frac{\lambda^{2}}{M^{2}}+\vec{X}\,^{2}+Y^{2}\bigg)P_{0}\Bigg]+\frac{1}{4P_{0}}\\ \nonumber
K_{i}&=\frac{1}{2}\bigg(K_{0}\frac{P_{i}}{P_{0}}+\frac{P_{i}}{P_{0}}K_{0}\bigg)-DX_{i}-X_{i}D
-\frac{P_{i}}{2P_{0}^{2}}=\\ \nonumber
&=\frac{1}{2}\bigg(K_{0}\frac{P_{i}}{P_{0}}+\frac{P_{i}}{P_{0}}K_{0}\bigg)+D\frac{1}{P_{0}}M_{0i}+M_{0i}\frac{1}{P_{0}}D
\end {align}
We see that essentially only special conformal generators must be, apart from simple symmetrization, supplemented with additional terms.
\par In order to construct an explicit form of the representation note that $P_{i}$ and $M$ form a maximal set of commuting observables. Therefore, they can be chosen as multiplication operators. The Hilbert space of states consists of functions $f(\vec{p},m)$ and the scalar product can be defined as
\setlength{\jot}{15pt}
\begin {align}
\label {al44}
(f,g)\equiv \int\limits^{\infty}\limits_{0} dm\int\frac{d^{3}\vec{p}}{p_{0}}\,\;\overline{f(\vec{p},m)} \;g\;(\vec{p},m)\quad \text {;}
\end {align}
here the factor $\sfrac{1}{p_{0}}$ is inserted to stress the relationship with Poincar\'{e} symmetry; have we changed it merely the explicit form of the representation would change.
\par The operators $\vec{X}$ and $Y$ can be taken as
\setlength{\jot}{15pt}
\begin {align}
\label {al45}
X_{i}\equiv i \frac{\partial}{\partial p_{i}}-i\frac{p_{i}}{2p_{0}^{\phantom{0}2}}\\ \nonumber
Y\equiv i \frac{\partial}{\partial m}-i\frac{m}{2p_{0}^{\phantom{0}2}}
\end {align}
It is worth noticing that, contrary to $\vec{X}$, $Y$ can be made only symmetric and not selfadjoint. Moreover, in order to make group generators selfadjoint some care must be exercised (see below).
\par It is convenient to change the variables from $\vec{p}$, $m$ to $\vec{p}$, $p_{0}\;(f(\vec{p},m)=\tilde{f}(\vec{p},\sqrt{p_{0}^{2}-\vec{p}\,^{2}}))$. Then the scalar product takes the form
\setlength{\jot}{15pt}
\begin {align}
\label {al46}
(\tilde{f},\tilde{g})=\int\limits_{v^{+}}\frac{d^{4}p}{(p^{2})^{\sfrac{1}{2}}}\;\overline{\tilde{f}(p)}\;\tilde{g}(p)
\end {align}
where $p^{2}\equiv p^{\phantom{0}2}_{0}-\vec{p}\,^{2}$,\;  $v_{+}=\{p\in M_{4}: p^{2}>0\;,\; p_{0}>0\}$.\\
Simultaneously,
\setlength{\jot}{15pt}
\begin {align}
\label {al47}
X_{i}&=i\frac{\partial}{\partial p_{i}}+i\frac{p_{i}}{p_{0}}\;\frac{\partial}{\partial p_{0}}-
i\frac{ p_{i}}{2p^{\phantom{0}2}_{0}}\\ \nonumber
Y&=i\frac{\sqrt{p^{2}}}{p_{0}}\;\frac{\partial}{\partial p_{0}}-i\frac{\sqrt{p^{2}}}{2p^{\phantom{0}2}_{0}}
\end {align}
Inserting the above expressions into eqs. (\ref{al43}) one finds the generators as differential operators of first or second order with coefficients being rational functions of $p_{\mu}$. They look slightly complicated and won't be written out explicitly here. However, we would like to show that our representation fits into the Mack classification. To see this note first that, when subduced to Poincar\'{e} subgroup, it decomposes into direct integral of irreducible representations of the latter, each corresponding to spin $s=0$ and mass $0<m<\infty$. Therefore, according to Ref. \cite{b34}, it corresponds to $j_{1}=j_{2}=0$. Accordingly, the transformation to Minkowski space is given by standard Fourier transform with no intertwining operators involved. In Mack's approach the generators acting in Minkowski spacetime take the form of first order differential operators with coefficients being the polynomials of at most second order. In momentum space this corresponds to the at most second order operators with linear (in momentum) coefficients. In order to put in this form the generators constructed by the method described above we make the following similarity transformation
\setlength{\jot}{15pt}
\begin {align}
\label {al48}
&f'(p)=(p^{2})^{-\gamma}\tilde{f}(p)\\ \nonumber
&\Gamma'\bigg(p,\frac{\partial}{\partial p}\bigg)=(p^{2})^{-\gamma}\,\Gamma\bigg(p,\frac{\partial}{\partial p}\bigg) (p^{2})^{\gamma}
\end {align}
with $\Gamma$ standing for any generator of conformal group. Putting 
\setlength{\jot}{15pt}
\begin {align}
\label {al49}
4\gamma^{2}-2\gamma=\lambda^{2}
\end {align}
we find that the transformed operators acquire the correct form. The scalar product is now
\setlength{\jot}{15pt}
\begin {align}
\label {al50}
(f',g')=\int\limits_{v_{+}}d^{4}p(p^{2})^{2\gamma-\frac{1}{2}}\;\overline{f'(p)}\,g'(p)
\end {align}
Eq. (\ref{al49}) has two real solutions
\setlength{\jot}{15pt}
\begin {align}
\label {al51}
\gamma = \frac{1}{4}(1\pm \sqrt{1+4\lambda^{2}})
\end {align}
Once the new form of momentum space generators is obtained one can take plain (as noted above, due to the absence of spin, no intertwiners are necessary) Fourier transform which yields
\setlength{\jot}{15pt}
\begin {align}
\label {al52}
P_{\mu}&=i\frac{\partial}{\partial x^{\mu}}\\ \nonumber
M_{\mu\nu}&=i\bigg(x_{\mu}\frac{\partial}{\partial x_{\nu}}-x_{\nu}\frac{\partial}{\partial x_{\mu}}\bigg)\\ \nonumber
D&=-ix^{\mu}\frac{\partial}{\partial x^{\mu}}+i(\delta-4)\\ \nonumber
K_{\mu}&=i\bigg(2x_{\mu}x^{\nu}\frac{\partial}{\partial x^{\nu}}-x^{2}\frac{\partial}{\partial x^{\mu}}+(8-2\delta)x_{\mu}\bigg)\\ \nonumber
\delta&=2\gamma+\frac{3}{2}
\end {align}
The above expressions, together with the scalar product (\ref{al50}), expressed in terms of $\delta$, coincide with the relevant formulae in Mack's paper (case (\ref{al3}) with $j_{1}=j_{2}=0$). There is, however, one obstacle. According to Mack, the lowest eigenvalue of the conformal Hamiltonian, $\delta$, obeys in this case $\delta>1$. By combining eqs. (\ref{al51}) and (\ref{al52}) one finds
\setlength{\jot}{15pt}
\begin {align}
\label {al53}
\delta_{\pm}=2 \pm\frac{1}{2}\sqrt{1+4\lambda^{2}}
\end {align}
For $\lambda^{2}$ running from $0$ to infinity $\delta_{+}\in (\frac{5}{2},\infty)$ and $\delta_{-}\in(-\infty,\frac{3}{2})$ while we should have $\delta\in(1,\infty)$. We do not fully understand this situation. We constructed the generators of conformal group as formally selfadjoint (i.e., in fact, symmetric) differential operators; it is understandable that by imposing more restrictive condition that they are selfadjoint in precise mathematical sense we further restrict the allowed values of $\delta$. This explains why we should reject the values $\delta \in (-\infty,1\rangle$. However, the interval $\langle\frac{3}{2},\frac{5}{2}\rangle$ seems to be excluded from our family while, according to Mack's construction, it is allowed.
 
\section{Conclusions and outlook} 
\label{IV}
\par We have shown how to construct the conformally invariant Hamiltonian dynamics in explicitly $SO(4,2)$ covariant way for nongeneric six- and eightdimensional orbits. In the former case one obtains massless fixed helicity particles while in the latter - spinless particles of varying mass. The resulting Hamiltonian systems can be canonically quantized. The ordering problems are here relatively mild. However, it appears that in the case of eightdimensional orbits some representations belonging to the family which is expected to correspond to those orbits cannot be recovered. The reason for that is unclear for us. 
\par Let us make some remarks concerning the interesting subclass of the irreducible representations of $SO(n,2)$ groups (here, $SO(4,2)$), called singletons. The notion of singletons dates back to Dirac \cite{b41} and Flato and Fronsdal \cite{b42} who considered the case $n=3$. A nice review of this class of representations, for general $n$, is given in Ref. \cite{b43}. Singletons can be defined in various ways. In particular, they may be viewed as the representations of $SO(n,2)$ which remain irreducible after subducing to $ISO(n-1,1)$ \cite{b43,b44}. In the present context, $n=4$, the latter are helicity $s$ representations of Poincare group $ISO(3,1)$. Therefore, the singleton representations are obtained by quantizing the Hamiltonian dynamics on sixdimensional (co)adjoint orbits of $SO(4,2)$.
\par It is interesting to consider in some detail the definition of the relevant orbits. To this end let us note that singletons can be related to the symmetry algebras of higher-spin dynamics. The latter, denoted by $\mathfrak{h}\mathfrak{s}_{d+2}$, may be obtained by taking the quotient of the $sO(d+1,2)$ universal enveloping algebra by an ideal vanishing on the singleton module \cite{b45,b46,b47}. For $d=n-1=3$ one obtains $\mathfrak{h}\mathfrak{s}_{5}$ algebra (actually, one-parameter family of algebras). The corresponding ideal is generated by two quadratic relations among generators of $SO(4,2)$ \cite{b48}. Their classical counterparts read 
\setlength{\jot}{15pt}
\begin {align}
\label {al54}
I_{ab}=\zeta_{a}^{\phantom{a}c}\zeta_{cb}+\frac{1}{6}C_{2}g_{ab}
\end {align}

\setlength{\jot}{15pt}
\begin {align}
\label {al55}
J_{ab}=\varepsilon_{abcdef}\zeta^{cd}\zeta^{ef}\mp 8\sqrt{\rho}\,\zeta_{ab}
\end {align}
Now, it is straightforward to check that the canonical points on coadjoint orbits, considered in Sec. II, obey (\ref{al55}) for any $\rho =\frac{1}{6}C_{2}\geq 0$. $J_{ab}$ is a tensor under coadjoint action so it vanishes on the whole orbit defined by eq. (\ref{al54}). Therefore, eq. (\ref{al55}) does not bring new information. This conclusion is no longer valid on the quantum level due to the ordering problems. The counterparts of the classical generators (\ref{al54}), (\ref{al55}) of the relevant ideals acquire additional terms and both generators are necessary \cite{b49,b50,b48}.
In the particular case $\rho=0$ we obtain the singleton representation called Rac. On the quantum level it describes massless particles with vanishing helicity. Its classical counterpart is a Hamiltonian dynamics on sixdimensional phase space. It has an interesting property of admitting global Darboux coordinates which transform as threevectors under the space rotations \cite {b36}. Let us note that neither Rac nor its classical counterpart contain any free parameter (this is clearly seen from the form of the canonical point defining the relevant orbit). 
\par More general analysis of coadjoint orbits of conformal group in four dimensions, the corresponding Hamiltonian systems and their quantization will be given elsewhere.
\par Finally, it is interesting to note that some nongeneric orbits of $SO(2,n)$ were considered in the context of $n-1$-dimensional Kepler problem; however, the choice of Darboux coordinates and, of course, the Hamiltonian were different \cite{b37,b38,b39,b40}.
\\
\\
{\bf Acknowledgments}
\\
We gratefully acknowledge inspiring discussions with Joanna Gonera, Katarzyna Bolonek-Laso\'n, Krzysztof Andrzejewski and Cezary Gonera. We are also very grateful to the anonymous reviewer who pointed out the connections between the subject of our paper and the symmetry algebras of higher-spin theories raising, in particular, the question concerning singleton representations.

\end {document}